\documentclass[final,5p,times,twocolumn]{elsarticle}

\usepackage{epsfig}

\usepackage{amssymb}
\journal{ }

\begin{document}

\begin{frontmatter}


%\title{aaa\tnoteref{label1}}
%\tnotetext[label1]{}

\ead{nozzoli@gmail.com}

\title{TeV dark matter in the disk}


\author{F. Nozzoli}

\address{
Dipartimento di Fisica, Universit\'a degli Studi di Roma ``Tor
Vergata'' \\ 
Via della Ricerca Scientifica 1, I-00133 Rome,
Italy}

\begin{abstract}
DAMA annual modulation data and, CoGeNT, CDMS-II, EDELWEISS-II, CRESST
excesses of events over the expected background are 
reanalyzed in terms of a dark matter particle signal 
considering
the case of a rotating halo.
It is found that DAMA data favor the configurations
of very high mass dark matter particles in a
corotating cold flux.
A similar high-mass/low-velocity solution would be 
compatible with the observed events in
CoGeNT, CDMS-II, EDELWEISS-II and CRESST experiments and could be of 
interest
in the light of the
positron/electron excess 
measured by Pamela and Fermi in cosmic rays.

\end{abstract}


\end{frontmatter}

% \linenumbers

\section{Introduction}
\label{intro}

Since 1996 the sodium iodide experiments of DAMA
collaboration (DAMA/NaI and DAMA/LIBRA) have 
measured an annual modulation of the single-hit 
counting rate which has the proper features expected
for a dark matter induced signal \cite{dama}.

More recently, other experiments (CoGeNT \cite{cogent}, 
CDMS-II \cite{cdms,cdms2}, EDELWEISS-II \cite{edw}, 
CRESST \cite{cresst})
have reported a preliminary observation of some excess of 
events relative to the expected backgrounds. 

The DAMA annual modulation signal and the other experiment
excesses, if interpreted as dark matter with dominant spin
independent interaction in the isothermal
halo model, implies that dark matter particles possess
a mass in the range of 5-15 GeV and an elastic scattering
cross section with nucleons in the order of $10^{-4}$ pb
\cite{weiner,bottino,bottino2}.

In this paper the same data are reanalyzed relaxing the
hypothesis of isothermal halo model, however it is assumed that
the dark matter local velocity distribution can 
still be approximated as a single Maxwellian flux:

\begin{equation}
f(\vec{v},\vec{v}_e) = \frac{1}
{\left(\pi v_0^2 \right)^{3/2}}
e^{-\left(\vec{v}+\vec{v}_e\right)^2/
v_0^2} .
\label{eq:fdv}
\end{equation}

Here the Earth velocity relative to the dark matter flux
is given by: $\vec{v}_e = \vec{v}_{\oplus}(t) + 
\vec{v}_{\odot}-\vec{v}_{DM}=\vec{v}_{\oplus}(t) +
\vec{v}^{LSR}_{\odot}+\vec{v}_{LSR}-\vec{v}_{DM}$; 
where: $\vec{v}_{\oplus}(t)$
is the Earth velocity in the solar system frame;
$\vec{v}^{LSR}_{\odot} = 
\vec{v}_{\odot}-\vec{v}_{LSR} =
(10.0,5.25,7.17)$ km/s
is the Sun velocity relative to the Local
Standard of Rest (LSR),
and $\vec{v}_{LSR} \simeq (0, 220 \pm 30, 0)$ km/s
\cite{mccabe}.
Therefore assuming a rotating dark matter 
halo ($\vec{v}_{DM} \neq 0$) one can write:  
$\vec{v}_{LSR}-\vec{v}_{DM} \simeq (0,v_{lag},0)$,
where $v_{lag}$ is the LSR velocity with respect
to the dark matter flux. 
Fixing $v_{lag} = v_0 \simeq 220$ km/s the 
eq. (\ref{eq:fdv}) provides the isothermal halo model,
however, in this analysis, the $v_0$ and $v_{lag}$ parameters
are kept free and
it is important to note that configurations
of $v_0$ and $v_{lag}$ that are far from the isothermal
halo ones can be physically meaningful\footnote{as an example 
$\Lambda$CDM halo simulations with baryons 
predict a corotating dark disk having 
$v_0 \sim 50$ km/s and $v_{lag} \sim 50$ km/s \cite{disk,disk2}}.

To avoid parameter proliferation, only the case
of dominant spin independent interaction for
elastically scattering dark matter will be considered
and the effects of uncertainties in the values adopted
for other parameters (quenching factor, form factor, 
possible presence of channeling, etc..)
will be neglected.
Therefore a four-parameter space $(v_0,v_{lag},M_W,\xi_0\sigma_p)$
will be considered here, where $M_W$ is the particle mass,
$\sigma_p$ is the proton cross section and $\xi_0 = 
\frac{\rho_{DM}}{0.3 GeV/cm^3}$ is the dark matter density in units of
$0.3 GeV/cm^3$.

\section{Experimental observables}
\label{data}

In this section the data used in the 
analysis are listed
for each 
experiment under
consideration:

\subsection{DAMA/NaI and DAMA/LIBRA}
The total exposure of 1.17 ton$\times$yr 
of NaI(Tl) provides three complementary observables:

\noindent
a) A modulated time behavior in the 2-6 keV window
(see data in fig. (\ref{fg:mod}) taken from fig. (4) 
of ref. \cite{dama})

\noindent
b)
The energy distribution of the observed modulation
amplitude, assuming a fixed phase $t_0=152.5$ d
(see data in fig. (\ref{fg:sm}) taken from fig. (6)
 of ref. \cite{dama}).
In the following analysis
the data in the 2-8 keV interval will be considered.

\noindent
c)
The energy distribution of the unmodulated
counting rate 
(see data in fig. (\ref{fg:s0}) taken from 
fig. (27) of ref. \cite{libra}).
This energy distribution provides a limit for 
the sum of
background and unmodulated dark matter induced signal
and therefore the limit
of 0.25 cpd/(kg $\times$ keV) 
for the possible unmodulated 
dark matter induced signal,
is cautiously assumed 
in the following analysis; 
this choice allows large
space for the presence of a low energy background component
in the measured counting rate.

\subsection{CoGeNT}

The data of fig. (3) of ref. \cite{cogent} are considered for the 
exposure of 330g $\times$ 56d collected by CoGeNT germanium
detector.

Only the data in the 0.4-0.9 keV window have been used
for the evaluation of the dark matter allowed configurations
in the hypothesis that this excess is induced by dark matter
elastic scattering; however
the whole energy interval in fig. (3) of ref. \cite{cogent}
 is considered for
the evaluation of the upper limit.
   
\subsection{CDMS-II and EDELWEISS-II}
The exposure of 969 kg $\times$ d collected by CDMS-II
germanium detectors \cite{cdms2} is considered.
Eleven events were observed within
the recoil acceptance region passing the rejection cuts
in the 10-150 keV energy range.
The neutron background is not able to explain
the CDMS-II measured events; however
some of these events
could be ascribed to surface background,
in particular for the low energy region. 
In the following the hypothesis that
the measured event excess could be due to 
dark matter elastic scattering is considered.
Moreover the data of the very low energy analysis
of CDMS-II (see fig. (1) of ref. \cite{cdmsle})
are 
also considered in
the evaluation of the upper limit. 
The recent result of EDELWEISS-II \cite{edw}
(where five recoil events are measured collecting the exposure
of 384 kg $\times$ d) seems to be compatible with the CDMS-II data;
therefore, for simplicity, only the CDMS-II data will be considered 
in this analysis.

\subsection{CRESST}
The preliminary exposure of 
564 kg $\times$ d collected by 
CRESST - $CaWO_4$ detectors is considered \cite{cresst}.
In the energy range $\sim 15 - 40$ keV (the lower threshold
is different for different detectors)
38 events are observed in the Tungsten 
recoil band and
52 events in the Oxygen one.
Despite the fact that $^{206}Pb$ recoils from $\alpha$ decay 
of $^{210}Po$ can contribute to the background
in the higher energy part of the Tungsten recoil band and
that the Oxygen recoil band is partially overlapped by
the $\alpha$ recoil band, only a fraction of the observed 
events can be ascribed to the evaluated background.
In this analysis, to account for the possible impact of
a confirmed excess in CRESST data, 
the case where 30 events of Tungsten recoil and 30 of
Oxygen recoil are induced by dark matter elastic scattering
will be considered;
this example will be generically addressed as CaWO$_4$.

\section{Parameter estimation}

The joint estimation of the four parameters
$(v_0,v_{lag},M_W,\xi_0\sigma_p)$ confidence interval
has been obtained by solving:

\begin{equation}
-2ln L(v_0,v_{lag},M_W,\xi_0\sigma_p)+2ln L_{max}
=\Delta
\label{eq:dl}
\end{equation}
for the appropriate values of $\Delta$
($\Delta = 7.78$ and 13.28 for 90\%
and 99\% C.L. respectively).
In the eq.(\ref{eq:dl}) 
$L(v_0,v_{lag},M_W,\xi_0\sigma_p)$ is the 
global likelihood function and 
$L_{max}$ is the likelihood maximum
value over the four parameter space.
In the estimation of confidence intervals,
the Gaussian approximation 
has been adopted for the likelihood
of DAMA/NaI + DAMA/LIBRA data;
moreover, the possible presence of 
unknown background explaining part
or all of the measured events for the other experiments
has been considered.
\begin{figure}[t]
\begin{center}
%\vspace{-0.5cm}
\mbox{\epsfig{figure=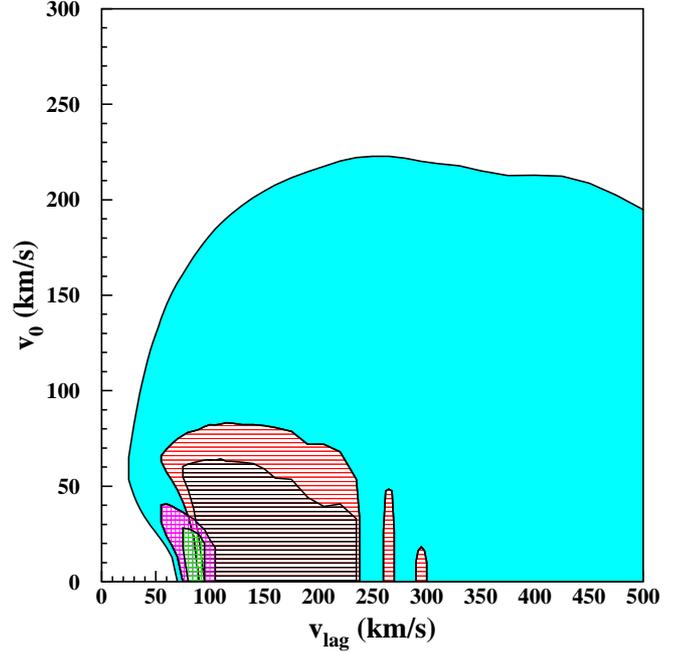,height=9.5cm}}
%\vspace{-0.3cm}
\caption{Horizontally hatched areas:
allowed configurations (90\% and 
99\% C.L.)
for unconstrained
DAMA/NaI + DAMA/LIBRA data. 
Cross hatched areas:
allowed configurations (90\% and 99\% C.L.)
for DAMA/NaI + DAMA/LIBRA data 
combined with CoGeNT, CDMS-II
and CRESST
data. 
Filled area: configurations having a C.L. better
than the one of isothermal halo model ($v_0 = 220$ km/s 
and 
$v_{lag} = 220$ km/s)  
for DAMA/NaI + DAMA/LIBRA data unconstrained.}
\label{fg:2d}
\end{center}
\end{figure}

In fig. (\ref{fg:2d}) the projection of the 
confidence interval surface (90\% and 99\% C.L.)
in the plane ($v_0$ vs $v_{lag}$) is shown for two cases:

\begin{itemize}
\item{ unconstrained DAMA/NaI + DAMA/LIBRA data 
(horizontally hatched 
area)}
\item{ DAMA/NaI + DAMA/LIBRA data combined with
CoGeNT, CDMS-II and CRESST data (cross-hatched area)}
\end{itemize}

It can be noted that DAMA/NaI+DAMA/LIBRA data 
favor configurations having low velocity dispersion
(low $v_0$) and relatively low $v_{lag}$ which
would imply a relatively cold and corotating dark matter
flux in the Galaxy.

Combining the DAMA data with the constraints from
other experiments further strengthens this indication.

As a comparison, in fig. (\ref{fg:2d}),
the projection of the configurations having a C.L. better
than the one of isothermal halo model ($v_0 = 220$ km/s
and $v_{lag} = 220$ km/s)
for unconstrained DAMA/NaI + DAMA/LIBRA data, is reported.

\begin{figure}[t]
\begin{center}
%\vspace{-1.cm}
\mbox{\epsfig{figure=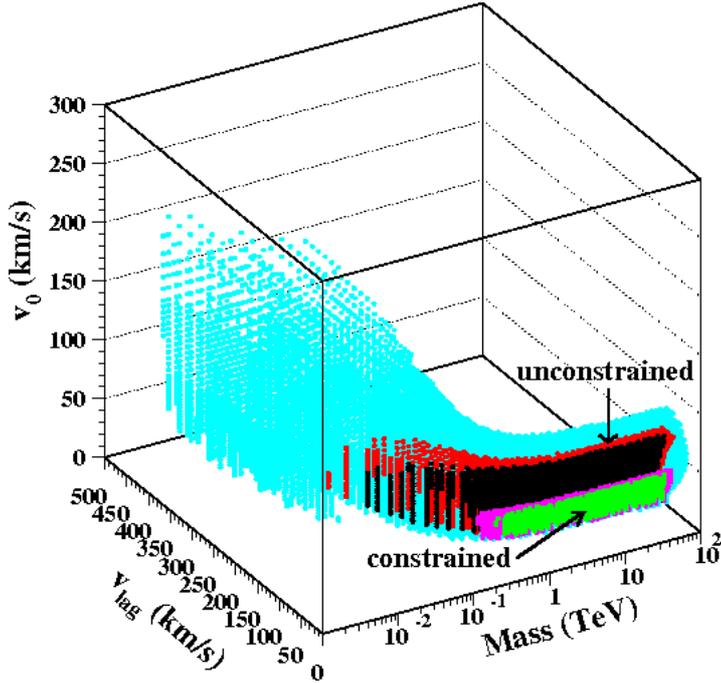,height=9.0cm}}
%\vspace{-0.3cm}
\caption{Allowed configurations in the volume
($v_0,v_{lag},M_W$).
The confidence levels and the cases of 
unconstrained/constrained DAMA data
adopt the same palette code 
used for fig. (\ref{fg:2d}).
The configurations of very heavy dark matter 
particles are favored. 
}
\label{fg:3d}
\end{center}
\end{figure}

In fig. (\ref{fg:3d}) the allowed configurations
in the volume ($v_0,v_{lag},M_W$) are shown.
The confidence levels and the cases of 
unconstrained/constrained DAMA data
adopt the same palette code 
used for fig. (\ref{fg:2d}).
It may be noted that configurations having
low $v_0$ and low $v_{lag}$ require a
very high $M_W$; only a similar configuration
would provide a flux of dark matter particles with
enough kinetic energy to allow
nuclear recoils events 
beyond the experimental thresholds.

\section{Comparison of the annual modulation signal
with respect to the case of isothermal halo model}

Here, the expected dark matter annual modulation 
signal features in NaI(Tl) are compared assuming
the four different models listed in table (\ref{tb:tb}).
\begin{table}[!hb]
\begin{center}
\resizebox{0.4\textwidth}{!}{
\begin{tabular}{|c|cccc|}
\hline
Model & $v_0$ (km/s) & $v_{lag}$ (km/s) & $M_W$ & $\xi_0 \sigma_p$ (pb)\\
\hline
\hline
a) & 220 & 220 & 60 GeV & $1.3 \times 10^{-5}$ \\
\hline
b) & 220 & 220 & 10 GeV & $9.3 \times 10^{-5}$ \\
\hline
c) & 10 & 95 & 90 TeV & $5.8 \times 10^{-4}$ \\
\hline
d) & 20 & 75 & 90 TeV & $4.6 \times 10^{-4}$ \\
\hline
\end{tabular}}
\caption{Models adopted in fig. (\ref{fg:mod}), (\ref{fg:sm})
and (\ref{fg:s0}).}
\label{tb:tb}
\vspace{-0.5cm}
\end{center}
\end{table}

\begin{figure}[!t]
\begin{center}
\mbox{\epsfig{figure=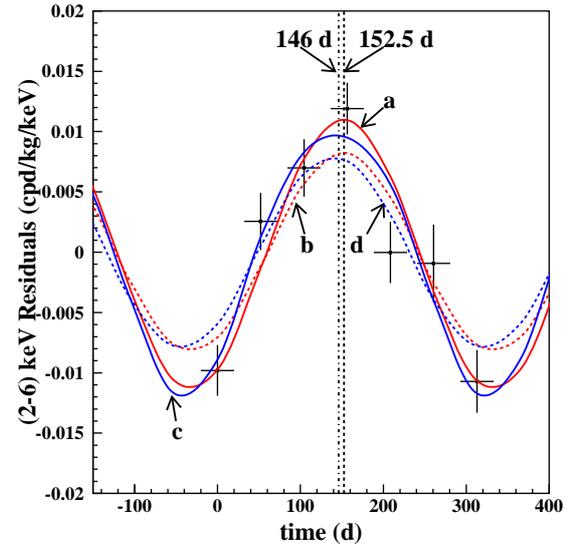,height=8.0cm}}
\caption{
Expected modulation behavior for NaI(Tl) in the 2-6 keV
region for the models listed in table (\ref{tb:tb}).
Data points are taken from fig. (4) of ref. \cite{dama}.
}
\label{fg:mod}
\end{center}
\end{figure}

In fig. (\ref{fg:mod}) the expected modulation
behavior in the 2-6 keV energy region of DAMA for the four 
considered models is shown.
The vertical lines mark the time 
of $152.5 d \sim 2^{nd}$ of June 
(where the maximum of the modulation amplitude
is expected for a non-rotating halo model)
and $146$ d where a maximum can be easily achieved, for example, 
assuming a corotating flux.
The data points are taken from fig. (4) of ref. \cite{dama} and 
represent
the annual modulation signal measured by DAMA/NaI + DAMA/LIBRA.
The measured time of maximum of the modulation in DAMA 
is $146 \pm 7$d, which is compatible 
both with non-rotating as well as with many of the rotating halo models. 
It is important to note that the modulation behavior
is roughly sinusoidal but for some extremal models
also large departures from a pure sinusoid can be found.
\begin{figure}[!tb]
\begin{center}
\mbox{\epsfig{figure=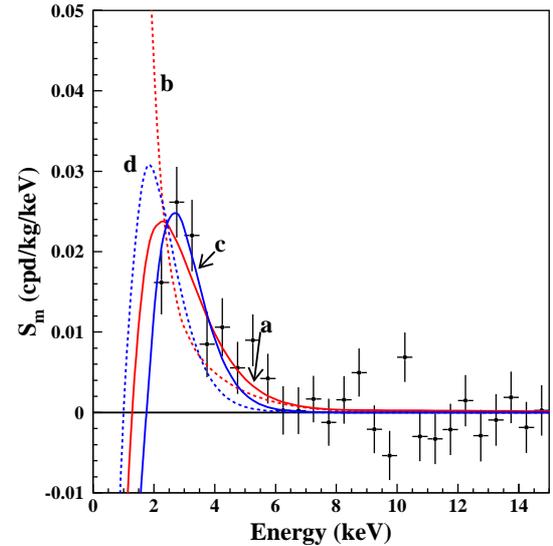,height=8.0cm}}
\caption{
Expected energy distribution of modulation amplitudes ($S_m$)
for the models  listed in table (\ref{tb:tb}).
Data points are taken from fig. (6) of ref. \cite{dama}.
}
\label{fg:sm}
\end{center}
\end{figure}
In fig. (\ref{fg:sm}) 
the expected energy distribution of modulation amplitudes ($S_m$)
for the four considered models is shown.
The data points are taken from fig. (6) of ref. \cite{dama} and 
represent
the annual modulation amplitude energy distribution measured by 
DAMA/NaI + DAMA/LIBRA.
\begin{figure}[!tb]
\begin{center}
\mbox{\epsfig{figure=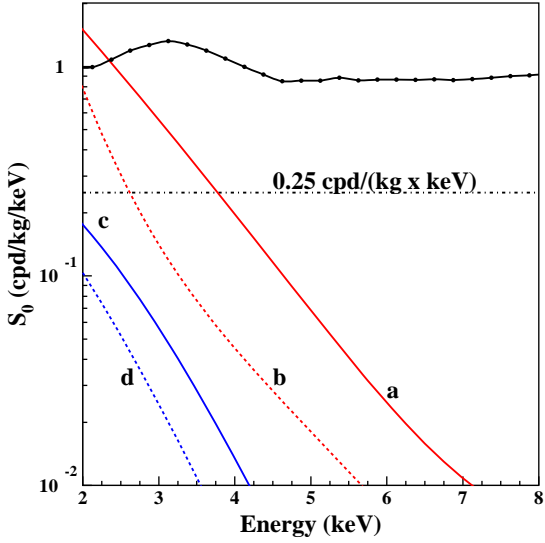,height=8.0cm}}
\caption{
Expected energy distribution of the unmodulated part of the 
counting rate ($S_0$)
for the four models  listed in table (\ref{tb:tb}).
Data points are taken from fig. (27) of ref. \cite{libra}.
}
\label{fg:s0}
\end{center}
\end{figure}
In fig. (\ref{fg:s0}) 
the expected energy distribution of the unmodulated part of the 
counting rate ($S_0$)
for the four considered models is shown.
The data points are taken from fig. (27) of ref. \cite{libra} and 
represent the
measured counting rate of DAMA/LIBRA; they are the sum of the 
background and of the possible dark matter signal.
The dot-dashed line marks the limit of 0.25 cpd/(kg $\times$ keV)
cautiously assumed for the maximum allowed $S_0$ value in this 
analysis. 
It is important to note that corotating halo models offer a large
$S_m/S_0$ ratio allowing the presence
of a reasonable background component also in the low energy part of DAMA 
data. 

\section{Allowed regions fixing the halo: an example.}

In this section, as an example, the halo model will 
be specified to fixed $v_0$ and $v_{lag}$ values. 
In this fixed halo model, the 
2$\sigma$ confidence intervals
in the ($M_W$ vs $\xi_0 \sigma_p$) plane are evaluated
considering configurations having $\Delta < 6.18$
with respect to the maximum likelihood of the 
considered halo model.
\begin{figure}[!htb]
\begin{center}
\mbox{\epsfig{figure=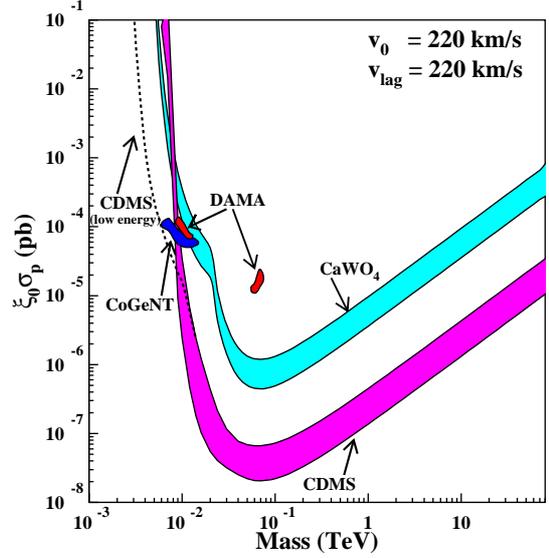,height=8.0cm}}
\caption{Allowed configurations at 2$\sigma$ C.L.
obtained for
the isothermal halo model with $v_0$ = 220 km/s and 
$v_{lag}$ = 220 km/s.}
\label{fg:220}
\end{center}
\end{figure}

In fig. (\ref{fg:220}) the allowed regions,
assuming the isothermal halo model 
($v_0 = v_{lag} = 220$ km/s) are shown.

\begin{figure}[!tb]
\begin{center}
\mbox{\epsfig{figure=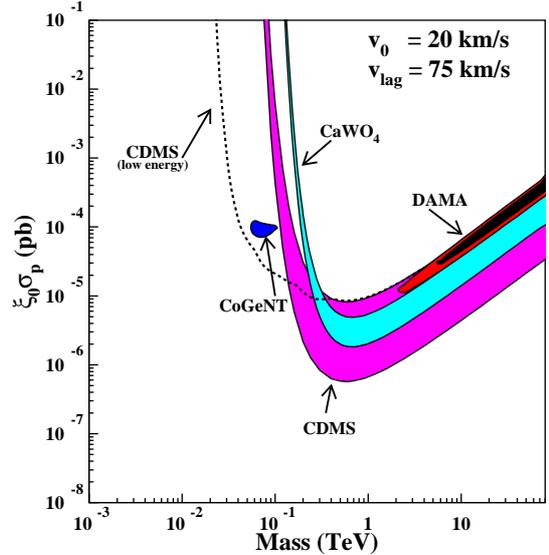,height=8.0cm}}
\caption{Allowed configurations at 2$\sigma$ C.L.
obtained for a cold corotating halo
 with $v_0$ =20 km/s and 
$v_{lag}$ = 75 km/s.
The black filled area inside the DAMA region marks the configurations
allowed at 1$\sigma$ C.L. ($\Delta <$ 2.3).
}
\label{fg:20_75}
\end{center}
\end{figure}

As a comparison, 
the allowed
regions for the case of a cold corotating
halo ($v_0$ = 20 km/s and $v_{lag}$ = 75 km/s) are given
in fig. (\ref{fg:20_75}). 
In both figures the dashed curve 
is the limit that can be evaluated
with CDMS-II when the low energy threshold data are
also considered \cite{cdmsle} .
It may be noted that compatibility among possible
positive hints for dark matter
could be achieved in models of 
very heavy particles ($M_W >$ few TeV)
forming a cold corotating halo.

As an example, for $v_0$ =20 km/s, $v_{lag}$ = 75 km/s, $M_W = 20$ TeV
and $\xi_0 \sigma_p = 10^{-4}$ pb, one would expect:

\begin{itemize}
\item{}only a fraction of 0.5\% of the total CoGeNT rate in the 0.4-0.9 
keV window due to dark matter elastic scattering
\item{}$\sim 13$ recoils measured in CDMS-II
\item{}$\sim 60$ recoils measured in  
CaWO$_4$ (CRESST-like) mainly 
expected to lie in
Tungsten band.
\end{itemize}

\section{Conclusions}
DAMA annual modulation data and, CoGeNT, CDMS-II, EDELWEISS-II,
CRESST excesses of events over the expected background have been 
reanalyzed in terms of a dark matter particle signal 
considering
the case of a rotating halo.
It has been found that the data favor the configurations
of very high mass dark matter particles in a
corotating cold flux.
This solution is intriguing since
$\Lambda$CDM halo simulations with baryons 
predict a corotating dark disk having 
$v_0 \sim 50$ km/s and $v_{lag} \sim 50$ km/s
\cite{disk,disk2}

A similar high-mass/low-velocity solution 
could be of interest
in the light of the
positron/electron excess 
measured by Pamela and Fermi in cosmic rays (see e.g. \cite{pam}).

Finally, the possibility of multi-component dark matter
should also be taken into account:
one could consider, for example, 
sterile neutrinos which form a dominant 
and warmer component of
dark matter in galaxies together
with a sub-dominant population of heavy WIMPs
mainly gathered in the core of
dwarf galaxies and, therefore, in the accreted dark disk.

%% The Appendices part is started with the command \appendix;%% appendix sections are then done as normal sections

%\section{}
%\label{}

%% References


\bibliographystyle{model1-num-names}
\bibliography{<your-bib-database>}



\end{document}